\documentclass{article}
\usepackage{latexsym,amsfonts,amssymb}

\begin{document}

\title{Structure of Multi-meron Knot Action}

\author{L. S. Isaev, A.P. Protogenov\footnote{e-mail: alprot@appl.sci-nnov.ru}  \\
\\
{\fontsize{10pt}{12pt}\selectfont
{\em 
Nizhny Novgorod State University,603950  Nizhny Novgorod 
}\/}\\
{\fontsize{10pt}{12pt}\selectfont
{\em 
Institute of Applied Physics of the RAS, 603950 Nizhny Novgorod
}}\/
}

\maketitle

\begin{abstract}
 We consider the structure of multi-meron knot action in the Yang-Mills 
 theory and in the $CP^1$ Ginzburg-Landau (GL) model. Self-dual equations 
 have been obtained without identifying orientations in the space-time 
 and in the color space. The dependence of the energy bounds  
 on topological parameters of coherent states in planar systems is also discussed.  
 In particular, it is shown that a characteristic size 
 of a knot in the Faddeev-Niemi model is determined by the Hopf invariant. 
\end{abstract}

\section{Introduction}
 Faddeev's  conjecture \cite{Fa} that the energy of the ground state and 
 properties of low-energy excitations in the non-Abelian field theory are 
 determined by topological invariants of knots and links has been recently 
 developed \cite{Fn} using the ${\bf n}$-field model. One of these 
 topological parameters is the Hopf invariant, which determines the knotting degree 
 of filamental manifolds, where the unit vector ${\bf n}$ is defined.

 The study of the behavior of a vortex filament tangle is an active 
 area of research in the field theory and attracts attention due to 
 several reasons \cite{Na}. The topological order associated with linking 
 at short distances exists against the background of disorder due to free motion of the tangle 
 filaments. Therefore, such systems of entangled filaments 
 contain data on their behavior in the ultraviolet, as well as infrared limit,   
 while systems of point particles do not contain.  
 In the present paper we discuss  some properties of field configurations 
 within the $SU(2)$ Yang-Mills theory and within the $CP^1$ GL model from this point of view.   
 
\section{The $SU(2)$ Yang-Mills theory}
 Let us consider the Langrangian density, 
 \begin{equation}
  L=\frac{1}{4g^2}Tr \hat{F}_{\mu\nu}^2,  
 \end{equation}
 \begin{equation}
  \hat{F}_{\mu\nu}=\partial_{\mu}\hat{A}_{\nu}-\partial_{\nu}
  \hat{A}_{\mu}+[\hat{A}_{\mu},\hat{A}_{\nu}],\,\,\,\,\hat{A}_{\mu}=A^a_
  {\mu}\frac{\sigma^{a}}{2i}.  
 \end{equation}
 We will use special anzatz \cite{Cho,Fn1} for the potential $A_{\mu}^a$: 
 \begin{equation} 
  A_{\mu}^a=\varepsilon^{abc}\partial_{\mu}n^{b}n^{c}(K+1). 
 \end{equation} 
 Here $g$ is the bare coupling constant, $K({\bf r},t)$ is the uknown scalar
 function, ${\bf n}$ is the unit vector field, and $\sigma^{a}$ denotes  
 Pauli matrices. Using Eq.(3), one can get
 \begin{eqnarray}
  \lefteqn{L=\frac{1}{2g^2}
  \left\{\delta_{\mu\nu}(\partial_{\lambda}{\bf n})^2-
  (\partial_{\mu}{\bf n})(\partial_{\nu}{\bf n})\right\}
  \partial^{\mu}K\partial^{\nu}K+} \\
  & & \qquad \frac{1}{4g^2}\left\{{\bf n}\cdot [\partial_{\mu}{\bf n},\partial_
             {\nu}{\bf n}]\right\}^2\left(K^2-1\right)^2 \nonumber.  
 \end{eqnarray}
 The Lagrangian density in the complete parametrization of the potential $A_{\mu}^a$
 was computed in Refs.\cite{Fn1} It is seen from Eq.(4) that multipliers in curly brackets 
 of the kinetic and potential parts of Eq.(4) play the role of the coupling constants 
 for the $K$-field and \emph{vice versa}. This takes place when ${\bf n}\not=
 {\bf r}/r$ and the foregoing soft version is applicable. Let us fix the ${\bf n}$-field 
 dynamics by the condition ${\bf n}=
 {\bf r}/r$. Then we will get a significantly simplified problem. Indeed, 
 in this case $A^a_{k}=\varepsilon^{akc}n^c(K+1)/r$ and 
 \[
  \left\{\delta_{\mu\nu}(\partial_{\lambda}{\bf n})^2-
  (\partial_{\mu}{\bf n})(\partial_{\nu}{\bf n})\right\}=
  \frac{\delta_{\mu\nu}}{r^2},\qquad ({\bf n}\cdot [\partial_{\mu}
  {\bf n},\partial_{\nu}{\bf n}])^2=\frac{1}{r^4}. 
 \]
 The Lagrangian density and the equation of motion for the scalar field $K(r,t)$
 takes the form
 \begin{equation} 
  L=\frac{1}{2g^2r^2}(\partial_{\mu}K)^2-\frac{1}{4g^2r^4}
    \left(K^2-1\right)^2 , 
 \end{equation} 

 \begin{equation}  
  r^2 \Box K=K(K^2-1), \,\, \, \, \Box=\frac{\partial^2}{\partial t^2}-\frac
 {\partial^2}{\partial r^2} . 
 \end{equation}
 The simplest solution \cite{FFD} is the meron configuration
 $
 K=t/\sqrt{t^2-r^2}.
 $
 Multi-meron configurations of the $K$-field were studied in detail in Ref.\cite{Pr}  
 Common property of the solutions of Eq.(6) is existence of 
 singularities of the field $K$ on closed surfaces located at a finite 
 distance from the point ${\bf r}=0$. Their existence makes it impossible to
 consider the infrared limit in the foregoing case, when degrees of freedom of 
 the ${\bf n}$-field are frozen.
 
 A fixed scalar degree of freedom, which is connected with the $K$-
 field can be obtained by integrating over this variable in the Feynman 
 integral. This problem was investigated in Refs.\cite{Fn1,Cho2,Sh,Gl}
 As a result, the main contribution to the model in Ref.\cite{Fa}  
 is characterized by the Lagrangian density
 \begin{equation}
  L=c_1(\partial_{\mu}{\bf n})^2+c_2\left({\bf n}\cdot [\partial_{\mu}
    {\bf n},\partial_{\nu}{\bf n}]\right)^2,  
 \end{equation}
 where $c_1$ and $c_2$ are effective coupling constants. The first term
 in Eq.(8) describes the infrared limit of the ${\bf n}$-field dynamics, 
 while the second determines the behavior at a short distance, where  
 topological effects of links are important.

 Now let us proceed to the general case. This can be done by considering the 
 self-dual equations using the anzatz in Eq.(3). Note that in this case of 
 low symmetry \cite{KR} these equations differ from those obtained for 
 spherically-symmetrical \cite{BPTSh,TH} or axially-symmetrical \cite{W} cases.
 Thus, we have
 \begin{equation}
  [\partial_{\tau}{\bf n},\partial_{k}{\bf n}]=
  \frac{1}{2}\varepsilon_{kms}[\partial_{m}{\bf n},\partial_{s}{\bf n}].
 \end{equation}
 \begin{equation}
  \partial_{\tau}K\,\partial_{k}{\bf n}-\partial_{k}K\,\partial_{\tau}{\bf n}=
  \frac{1}{2}\varepsilon_{kms}\left\{
  \partial_{m}K\,\partial_{s}{\bf n}-\partial_{s}K\,\partial_{m}{\bf n}\right\},
 \end{equation}
 where $\tau=it$ is the Euclidean time. Since the homotopic group $\pi_4(SU
 (2))=\mathbb Z_2$ is non-trivial \cite{W1}, we hope that there exists at
 least one non-trivial class of configurations 
 of fields $K$ and ${\bf n}$. 
 
 \section{The $CP^1$ Ginsburg-Landau model}
 
 We will use the GL functional, 
 \begin{eqnarray} 
  \lefteqn{F=\int d^{3}x\,\biggl[\sum_{\alpha}\frac{1}{2m}\left|\left(\hbar
           \partial_{k}+i\frac{2e}{c}A_{k}\right)\Psi_{\alpha}\right|^2+}\\
  & & \qquad \sum_{\alpha}\left(-b_{\alpha}|\Psi_{\alpha }|^{2}+\frac{c_
  {\alpha}}{2}|\Psi_{\alpha}|^4\right)+\frac{\bf B^{2}}{8\pi}\biggr] , 
  \nonumber
 \end{eqnarray}
 with a two-component order parameter
 $
  \Psi_\alpha=\sqrt{2m}\,\rho\,\chi_\alpha,\,\,\,\,\chi_\alpha=|\chi_\alpha
  |e^{i\varphi_\alpha}\,, 
 $                                                                             
 which satisfies the $CP^1$ condition, $|\chi_{1}|^{2}+|\chi_{2}|^{2}=1$. 
 In the (3+0)-dimensional case, Eq.(10) has the meaning of free energy and 
 $\Psi_{\alpha}$ is a two-component order parameter, which is used either 
 in the context of two-gap supercondutivity \cite{BFN} or in the 
 non-Abelian field theory as a Higgs doublet \cite{Ch3}. In (2+1)D 
 systems, Eq.(10) has the meaning of action, and the order 
 parameter $\Psi_{\alpha}$ gives a two-dimensional non-Abelian representation 
 of the braid group,   
 which is a specific group of symmetry for planar systems at permutations  
 of particles inside them.

 As was shown in Ref.\cite{BFN}, there exists exact mapping of  
 model (10) into the following version of the ${\bf n}$-field model: 
 \begin{eqnarray}
  \lefteqn{F=} \\ \nonumber
  & & \; \int d^{3}x\left[\frac{1}{4}\rho^{2}\left(\partial_{k}{\bf n}
         \right)^{2}+\left(\partial_{k}\rho \right)^{2}+\frac{1}{16}\rho^{2}
         {\bf c}^{2}+\left(F_{ik}-H_{ik}\right)^{2}+V(\rho, n_{3})\right].
 \end{eqnarray}
 The equation (11) was obtained with the use of  gauge invariant order parameter fields 
 of the unit vector $n^a={\bar \chi}\sigma^{a}\chi$, where $\bar\chi=
 (\chi_{1}^{\ast}, \chi_{2}^{\ast})$ and the velocity ${\bf c}={\bf J}/\rho^{2}$. 
 The total current, ${\bf J}=
 2\rho^{2}({\bf j}-4{\bf A})$, has paramagnetic $\left({\bf j}=i[\chi_{1}
 \nabla\chi_{1}^{\ast}-c.c.+(1 \to 2)]\right)$ and diamagnetic $(-4{\bf A})$ 
 parts. Besides, we used in Eq.(11) the following notations:
 $F_{ik}=\partial_{i}c_k-\partial_{k}c_i$, and $H_{ik}={\bf n}\cdot[\partial_{i}
 {\bf n}\times\partial_{k}{\bf n}]:=\partial_{i}a_{k}-\partial_{k}a_{i}$ and
 the dimensionless units: $L=(\xi_1+\xi_2)/2$ as the unit of length, 
 $\xi_{\alpha}=\hbar /\sqrt{2m b_{\alpha}}$ as the coherence
 length; $\hbar/L$ as the unit of {\bf c};   
 $c^2/(512\pi e^2 L^{2})$ as the unit of $\rho^2$ and, finally,   
 $\gamma /L$ with $\gamma=\left(c\hbar /e \right)^2/512\pi$ as the unit of energy.  

 Let us now enumerate some non-trivial (${\bf n}\not=const$) situations  
 that follow from Eq.(11) as a result of competition of the order parameters 
 $\rho$, ${\bf n}$, and ${\bf c}$: (i) ${\bf c}=0$, $\rho=const$; (ii) 
 ${\bf c}=0$, $\rho\not=const$; (iii) ${\bf c}\not=0$, $\rho=const$; (iv) 
 ${\bf c}\not=0$, $\rho\not=const$. In the case when ${\bf c}\not=0$, 
 $\rho\not=const$, and ${\bf n}=const$ we have the problem of a vortex state 
 structure in the classical GL model.

 In the limit (i) we get Eq.(7) \cite{Fn}. A numerical study of  
 knotted configurations of the ${\bf n}$-field in this model was performed  
 in Refs. \cite{GH,BS,HS}. The lower energy bound in this case, 
 \begin{equation}
  F \geqslant 32\pi^{2}\,|Q|^{3/4} , 
 \end{equation}
 is determined \cite{VK,KR,Wo} by the Hopf invariant, 
 \begin{equation}
  Q=\frac{1}{16\pi^{2}}\int d^{3}x\,\varepsilon_{ikl}a_{i}
    \partial_{k}a_{l}\,.
 \end{equation}
 If the compactification $\mathbb R^{3} \to S^{3}$ is used and ${\bf n} \in S^{2}$, then 
 the integer $Q\in\pi_{3}(S^{2})=\mathbb Z$ shows the degree of linking or 
 knotting of filamental manifolds, where the vector field ${\bf n}(x,y,z)$ 
 is defined. In particular, for two linked rings (Hopf linking) $Q=1$, for 
 the trefoil knot $Q=6$, etc. It is important to note that  
 $\pi_{3}(CP^{M})=0$ at $M>1$ and $\pi_{3}(CP^{1})=\pi_{3}(S^{2})=\mathbb Z$ 
 \cite{AW}. In the latter case, the order parameter is two-component
 \cite{BFN}, and linked or knotted soliton configurations are labeled by the
 Hopf invariant in Eq.(13).
 
 The characteristic size $R_Q$ of a knot can be found using Eq.(12) and the 
 minimum value of the free energy $F_{min}=2\sqrt{c_1c_2}$ for the radius
 $R_{min}=\sqrt{c_2/c_1}$. Therefore,  
 \begin{equation} 
  R_Q=\frac{c_2}{16\pi^2\left|Q\right|^{3/4}}. 
 \end{equation}
 We see that the characteristic size of such a knot is determined by a  
 combination of dynamical ($c_2$) and topological ($Q$) features of the
 system. In case (ii), we have the problem of a soft version of the 
 ${\bf n}$-field model \cite{Ni}. Case (iv) is most general. 

 We will focus now on limit (iii). Let us assume that
 $\rho$ can be found from the minimum value of the potential $V(\rho)$ and 
 the velocity ${\bf c}$ is not equal to zero. Eq.(11) in this case takes the
 following form: 
 \begin{eqnarray}
  \lefteqn{F=F_{n}+F_{c}-F_{int}=} \\ \nonumber
  & & \; \int d^{3}x\left[\left(\left(\partial_{k}
         {\bf n}\right)^{2}+H_{ik}^{2}\right)+\left(\frac{1}{4}{\bf c}^{2}+
         F_{ik}^{2}\right)-2F_{ik}H_{ik}\right]\,.
 \end{eqnarray}
 
 It is seen from Eq.(15) that the superconducting state with ${\bf c}\not=0$ 
 has the energy, which is less than the minimum in the case (i) due to 
 renormalization of the coefficient $(=1)$ in the second term of the 
 functional $F_{n}$. In order to find the lower free energy bound in the 
 superconducting state with ${\bf c}\not=0$, we will use the following 
 inequality \cite{PV}: 
 \begin{equation}
  F_{n}^{5/6}\,F_{c}^{1/2} \geqslant (32\pi^{2})^{4/3}\,|L| \, \, , 
 \end{equation}
 where
 \begin{equation}
  L=\frac{1}{16\pi^{2}}\int d^{3}x\,\varepsilon_{ikl}c_{i}\partial_{k}a_{l} 
 \end{equation} 
 is the degree of mutual linking \cite{AK,M} of the velocity ${\bf c}$ 
 lines and of the magnetic field ${\bf H} = [\nabla \times {\bf a}]$ lines. 
 It is also an integral of motion. \cite{M,ZK}

 Using the Schwartz-Cauchy-Bunyakovsky inequality, one will find  
 \cite{PV} that
 $ 
  F\geqslant F_{min}=(F_n^{1/2}-F_c^{1/2})^2.
 $ 
 Using Eq.(16) and the condition $Q\not=0$ the last inequality can be rewritten in the
 equivalent form
 \begin{equation}
  F\geqslant 32\pi^{2}\,|Q|^{3/4}\,(1-|L|/|Q|)^{2}\,\,.  
 \end{equation}
 The minimum value of free energy in Eqs.(18) corresponds to the 
 field configurations with $L=Q\not=0$ or ${\bf a}={\bf c}$. 
 They satisfy the self-dual condition $F_n=F_c$.

 It is clear from Eq.(18) that for all numbers $L<Q$ the energy of the 
 ground state is less than that described in the model \cite{Fn}, for which
 the inequality (12) is valid. The origin of the energy decrease can be 
 easily understood. Even under the conditions of existence of the 
 paramagnetic part ${\bf j}$ of the current ${\bf J}$, the diamagnetic 
 interaction in the superconducting state cancels the own current energy and
 a part of the energy related to the ${\bf n}$-field dynamics for all 
 state classes with $L<Q$. It is also important that the current (total  
 momentum of a superconducting pair) is non-zero in the superconducting 
 state. In this regard, the considered inhomogeneous state 
 with the current is similar to the state \cite{L} investigated in Refs.
 \cite{LO,FF}. 

\section{Discussion}
 Up to now the vector ${\bf A}$ has characterized the internal charge $U(1)$ gauge symmetry. 
 If we apply the external electromagnetic field, then the vector potential ${\bf A}$ will be equal to 
 the sum of internal and external gauge potentials. As a result, the 
 velocity ${\bf c}$ decreases due to the diamagnetism of the superconducting 
 state. This leads to suppression of the superconducting gap. The answer to the
 question on existence of full or partial Meisner screening in these states depends 
 on the result of the competition between contributions from neutral ${\bf j}$ and 
 charged $-4{\bf A}$ parts to the total current ${\bf J}$. 
 
 In the $(3+0)D$ case of free energy (10), Hopf invariant (13) is analogous 
 to the Chern-Simons action 
 $(k/4\pi)\int dt\,d^{2}x\,\varepsilon_{\mu\nu\lambda}a_{\mu}\partial_{\nu}a_{\lambda}$, 
 which determines strong correlations of $(2+1)D$ states \cite{Pr2} for $k \simeq 2 $. 
 In planar systems, this coefficient has the sense of braiding degree of excitation world 
 lines. In particular, for the semion $k=2$. Keeping in mind the relation of spatial 
 dimensionality of the systems in their quantum and statistical descriptions, 
 we note that the $(2+1)D$ dynamical case $k=2$ of open world line ends of 
 excitations is equivalent to the compact $(3+0)D$ statistical example of the Hopf linking  $Q=1$.

In conclusion, we have considered the structure of the multi-meron knot 
action and found the self-dual equation. 
Using the topological invariants of knots for the analysis of coherent phase states, 
we have also determined the conditions of applying the theory significantly 
based on the dynamics and topology of the ${\bf n}$-field model. 

\section*{Acknowledgements}
We are very grateful to A.~G. Abanov, S. A. Brazovskii, L.~D. Faddeev, A.~I. Larkin, Y.-S. Wu and G.~E. Volovik for useful discussions. One of the authors (AP) thanks 
Prof. M. Boiti for kind hospitality. This work was partially supported by the RFBR 
under Grant \# 01-02-17225.


\begin{thebibliography}{0}
\bibitem{Fa} L. D. Faddeev. Quantization of solitons. Preprint IAS-75-QS70, 1975.
\bibitem{Fn} L. D. Faddeev and A. Niemi, {\it Nature} {\bf 387}, 58 (1997).
\bibitem{Na} V. Katritch {\it et al.}, {\it Nature} {\bf 384}, 142 (1996).
\bibitem{Cho} Y. M. Cho. {\it Phys. Rev. Lett.} {\bf 46}, 302 (1981).
\bibitem{Fn1} L. D. Faddeev, A. Niemi, {\it Phys. Rev. Lett.} 
{\bf 82}, 1624 (1999); {\it Phys. Lett.} {\bf B449}, 214 (1999); {\bf 464}, 46 (1999). 
\bibitem{FFD} V. De Alfaro, S. Fubini and G. Furlan, {\it Phys. Lett.} {\bf B65}, 163 (1976).
\bibitem{Pr} A. P. Protogenov, {\it Phys. Lett.} {\bf B87}, 80 (1979).
\bibitem{Cho2} W.~S. Bae, Y.~M. Cho and S.~W. Kimm, hep-th/0105163.
\bibitem{Sh} S. Shabanov, hep-th/0004135. 
\bibitem{Gl} H. Gies, {\it Phys. Rev.} D {\bf 63}, 125023 (2001).
\bibitem{KR} A. Kundu and Y.~P. Rubakov, {\it J. Phys} {\bf A15}, 269 (1982).
\bibitem{BPTSh} A.~A. Belavin, A.~M. Polyakov, A.~S. Shwartz and Y. Tyupkin, 
{\it Phys. Lett.} {\bf B59}, 85 (1975).
\bibitem{TH} G. t'Hooft, {\it Phys. Lett.} {\bf D14}, 3432 (1976).
\bibitem{W} E. Witten, {\it Phys. Rev. Lett.} {\bf 38}, 121 (1977).
\bibitem{W1} E. Witten, {\it Phys. Lett.} {\bf B117}, 324 (1982).
\bibitem{BFN} E. Babaev, L.~D. Faddeev, A.~J. Niemi, {\it Phys. Rev.} B {\bf 65}, 100512 (2002).
\bibitem{Ch3} Y.~M. Cho, cond-mat/0112498.
\bibitem{GH} Y. Gladikowski and M. Hellmund, {\it Phys. Rev.} D {\bf 56}, 5194 (1997).
\bibitem{BS} R.~A. Battye and P.~M. Sutcliffe, {\it Phys. Rev. Lett.} {\bf 81}, 4798 (1998).
\bibitem{HS} J. Hietarinta and P. Salo, {\it Phys. Lett.} {\bf B451}, 60 (1999).
\bibitem{VK} A.~F. Vakulenko and L.~V. Kapitansky, {\it Sov. Phys. Dokl.} {\bf 24} 433 (1979). 
\bibitem{Wo} R.~S. Ward, {\it Nonlinearity} {\bf 12}, 1 (1999).
\bibitem{AW} A.~G. Abanov and P.~W. Wiegmann, hep-th/0105213.
\bibitem{Ni} M. L\"ubke, S.~M. Nasir, A. Niemi and K. Torokoff, hep-th/0106102.
\bibitem{PV} A.~P. Protogenov and V.~A. Verbus, {\it JETP Lett.} {\bf 76}, 53 (2002).  
\bibitem{AK} V.~I. Arnold and B.~A. Khesin. Topological methods in hydrodynamics. 
{\it Appl. Math. Sci.} {\bf 125}, chapt. 3.
\bibitem{M} A.~K. Moffatt, {\it J. Fluid Mech.} {\bf 106}, 117 (1969).
\bibitem{ZK} V.~E. Zakharov and E.~A. Kuznetsov, {\it Phys. Usp.} {\bf 40}, 1087 (1997).
\bibitem{L} A.~I. Larkin, (personal communication).
\bibitem{LO} A. I. Larkin and Yu. N. Ovchinnikov, {\it Sov. Phys. JETP} {\bf 20}, 762 (1965). 
\bibitem{FF} P. Fulde and R. A. Ferrell, {\it Phys. Rev.} A {\bf 135}, 550 (1964).  
%\bibitem{Pr1} A.~P. Protogenov, cond-mat/0205133.
\bibitem{Pr2} A.~P. Protogenov, {\it JETP Lett.} {\bf 73}, 255 (2001).
\end{thebibliography}
\end{document}